\documentclass[twocolumn,showpacs,amsmath,amssymb,floatfix,pre,a4]{revtex4}
\usepackage{graphicx}
\begin{document}
\title{Particle-wall collision statistics in the open circular billiard}
\author{J\"urgen F. Stilck}
\email{jstilck@if.uff.br}
\affiliation{Instituto de F\'{\i}sica\\
Universidade Federal Fluminense\\
Av. Litor\^anea s/n\\
24210-346 - Niter\'oi, RJ\\
Brazil}
\date{\today}

\begin{abstract}
In the open circular billiard particles are placed initially with a
uniform distribution in their positions inside a planar
circular vesicle. They all have velocities of the same magnitude,
whose initial directions are also uniformly distributed. No
particle-particle interactions are included, only specular elastic
collisions of the particles with the wall of the vesicle. The particles
may escape through an aperture with an angle $2\delta$. The
collisions of the particles with the wall are characterized by the
angular position and the angle of incidence. We study the evolution of
the system considering the probability distributions of these variables
at successive times $n$ the particle reaches the border of the
vesicle. These distributions are calculated analytically and measured in
numerical simulations. For finite apertures $\delta<\pi/2$, a
particular set of initial conditions exists for which the particles
are in periodic orbits and never escape the vesicle. This set is of
zero measure, but the selection of angular momenta close to 
these orbits is observed after some collisions, and thus the
distributions of probability have a structure formed by peaks. We
calculate the marginal distributions   
up to $n=4$, but for $\delta>\pi/2$ a solution is found for arbitrary
$n$. The escape probability as a function of $n^{-1}$ decays with an
exponent 4 for $\delta>\pi/2$ and evidences for a power law decay are
found for lower apertures as well.
\end{abstract}


\maketitle

\section{Introduction}
\label{intro}
In the theory of dynamical systems,  the study of the decay of simple  
hamiltonian systems \cite{bb90}-\cite{vk01} is of much interest. These
studies 
are an extension of earlier research concerning the closed version of such
systems, where the main question is their ergodicity. The pioneering
work in this area is centered on the system called Sinai billiard,
a circular billiard with a smaller circular exclusion area in
its interior \cite{s70}, which was shown to be ergodic. Other
two-dimensional ballistic billiards are known to be ergodic as well,
such as the Bunimowitch stadium \cite{b74,agh96}. One main interest in
the studies of open billiards is the decay dynamics for long
times. Bauer and Bertsch \cite{bb90} found an exponential decay in an
chaotic dynamics and a power law decay for a system with regular
dynamics. These first results concerning the integrable case where
later questioned by Legrand and Sornette \cite{ls90}, but 
it became clear that the difficulty in settling this question using
numerical experiments is
related to the high sensitivity of the results to initial conditions
\cite{bb91}. In a more detailed simulational study for the chaotic
two-dimensional  
Bunimovich stadium \cite{agh96}, algebraic tails were found at
sufficiently long times, but the weight of the algebraic tail tends to
zero in the limit where the size of the aperture vanishes.

In the classical circular billiard the (non-interacting) particles undergo
elastic specular collisions with the wall. Since two quantities are
conserved (kinetic energy and angular momentum), the system is
integrable and therefore non-ergodic \cite{rr99}. A recent study of
the open version of this system was undertaken by Vicentini and
Kokshenev \cite{vk01}, and the algebraic long time of the survival
probability was studied in detail, in the limit of a very small
opening ({\em weakly open billiard}), using a coarse-grained
description of the system. Here we study the problem making use of a
probabilistic approach: starting from a random initial condition of
the particles inside the vesicle, we calculate the joint probability
distribution (in the angular position of point where the border is
reached and the angle
of incidence) at successive times the border of the vesicle is
reached. This approach is not limited to small openings, actually it
allows a full analytical treatment when the opening $\delta$ is larger than
$\pi/2$ ($\delta=\pi/2$ corresponds to a semi-circular vesicle). It
will become clear below that for higher number of collisions with the
wall, those particles which are close to periodic orbits which will
never leave the vesicle will still survive, and thus a selection of
these orbits will occur with increasing number of collision $n$. This
selection of orbits according to the incidence angle (or,
equivalently, to the angular momentum) was already found in the
numerical experiments performed in \cite{vk01}, and emerges in a
simple way in the probabilistic calculations below. The algebraic
decay of the probability that the particle leaves the vesicle at the
$n$'th collision is obtained analytically for large openings
($\delta>\pi/2$). 

Let us define the problem. The initial position of a particle is
distributed uniformly inside the circular vesicle of radius $r_0$. The
particle has a velocity of modulus $v_0$, whose direction is also
uniformly distributed. After some time, this particle reaches the
border of the vesicle for the first time, at an angle $\theta_1$. If
$|\theta_1|<\delta$, the opening angle, the particle leaves the
vesicle, otherwise it collides with the wall and performs the second
straight part of its movement inside the billiard. If we use $r_0$ as
our unit of length and $r_0/v_0$ as our unit of time we may reduce the
problem to the case of unitary radius and modulus of velocity. 

In Sec. \ref{def} we define the problem in more detail and present the
analytical approach we adopted. The numerical study of the problem is
described in Sec. \ref{simu}, and the simulational results are
compared with the analytical ones and discussed. Further discussions
and the conclusion may be found in Sec. \ref{con}.

\section{Definition of the problem and analytical approach}
\label{def}
\begin{figure}
\begin{center}
\includegraphics[height=6.0cm]{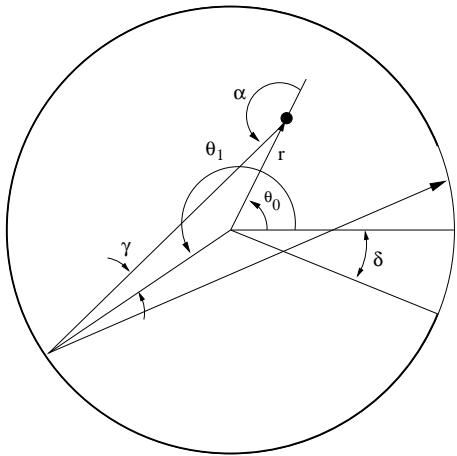}
\caption{Movement of a particle, depicted at its initial position. This
particle leaves the vesicle when it reaches the border for the second
time. The indicated variables are described in the text.}
\label{f1}
\end{center}
\end{figure}

As discussed in Sec. \ref{intro}, we consider a particle inside a
circular vesicle of unitary radius. The 
probability of the initial position, described by a radius $r$ ($0 \leq r \leq
1$) and an angle $\theta_0$ ($-\pi \leq \theta_0 < \pi$ ), is uniformly
distributed. The modulus of 
the initial velocity is unitary and the angle $\alpha$ ($-\pi \leq \alpha
<\pi$)between the 
position vector and the velocity is also uniformly distributed. Somewhat
later, the particle reaches the border of the vesicle for the first time. If
the 
angle $\theta_1$ where this occurs is such that $-\delta \leq \theta_1
\leq \delta$, the particle leaves the vesicle. Otherwise, the particle
collides in a specular way with the border of the vesicle, continuing its
movement until the border is reached for a second time at an angle
$\theta_2$. In figure \ref{f1} the trajectory of a particle is illustrated.

The movement of the particle is deterministic, but the initial conditions are
randomly distributed, and we are interested in the joint distribution of
probabilities at the $n$'th time the border is reached. The initial joint
distribution function is given by:
\begin{equation}
\rho_0(r,\alpha,\theta_0)=\frac{r}{2\pi^2},
\label{rho0}
\end{equation}
and this leads to the marginal distributions $\rho_0(r)=2r$ and
$\rho_0(\alpha)=\rho_0(\theta)=1/2\pi$, which are consistent with the initial
conditions stated above. Simple geometrical considerations allow us now to
relate the variables  $\gamma_1$ and $t_1$ and $\theta_1$ at the first time the
particle reaches the border with the original variables $r$, $\alpha$, and
$\theta_0$. Notice that $t_1$, the time of the first arrival at the border, is
numerically equal to the distance between the initial position and the
position of first arrival at the border. The angle of incidence $\gamma_1$ is
defined in the interval $-\pi/2 \leq \gamma_1 \leq \pi/2$, and negative values
correspond to particles moving clockwise in the vesicle. With this convention,
the signs of $\alpha$ and $\gamma$ are the same. The relations between the
variables may be found applying trigonometric considerations to the
initial segment of the trajectory illustrated in figure \ref{f1}. The
result is:
\begin{subequations}
\label{transf1}
\begin{eqnarray}
r &=& \sqrt{t_1^2-2t_1\cos(\gamma_1)+1} \\
\tan (\alpha) &=& \frac{\sin(\gamma_1)}{\cos(\gamma_1)-t_1} \\
\theta_0 &=& \theta_1+\alpha -\gamma_1
\end{eqnarray}
\end{subequations}
The joint distribution at the first arrival at the border is related to the
initial joint distribution as follows:
\begin{equation}
\rho_1(\gamma_1,t_1,\theta_1)=\rho_0(r,\alpha,\theta_0)
\left|\frac{\partial(r,\alpha,\theta_0)}{\partial(\gamma_1,t_1,\theta_1)}
\right|.
\end{equation}
The jacobian may then be calculated using expressions \ref{transf1}. The
result is: 
\begin{equation}
|J|=\frac{\cos(\gamma_1)}{\sqrt{t_1^2-2t_1\cos(\gamma_1)+1}},
\end{equation}
and using the initial distribution \ref{rho0}, we obtain the joint
distribution at the first arrival at the border
\begin{equation}
\rho_1(\gamma_1,t_1,\theta_1)=\frac{\cos (\gamma_1)}{2\pi^2}.
\label{rho1}
\end{equation}

From the joint distribution \ref{rho1} the marginal distributions may then be
calculated. The result for $t_1$ is:
\begin{eqnarray}
\rho_1(t_1) &=& \frac{1}{2\pi^2}\int_{-\pi}^{\pi}d\theta_1 
\int_{-\arccos(t_1/2)}^{\arccos(t_1/2)} \cos(\gamma_1) \; d\gamma_1
\nonumber \\
&=& \frac{2}{\pi}\sqrt{1-t_1^2/4}
\label{t1}
\end{eqnarray}
and the marginal distributions of the other two variables follow in a similar
way: 
\begin{eqnarray}
\rho_1(\gamma_1) &=& \frac{2}{\pi}\cos^2(\gamma_1), \label{g1}\\
\rho_1(\theta_1) &=& \frac{1}{2\pi}. \label{th1}
\end{eqnarray}
As expected, the distribution of $\theta_1$ is uniform. The probability that
the particle leaves the vesicle at the first arrival at the border is given
by: 
\begin{equation}
p_1=\int_{-\delta}^{\delta} \rho_1(\theta_1) \; d\theta_1=
\frac{\delta}{\pi}.
\label{p1}
\end{equation}

Due to angular momentum conservation, the incidence angle will be the same at
each subsequent time the particle reaches the border. Also, the time interval
between the first and $n$'th times at the border will simply be
$t^\prime_n=t_n-t_1=2(n-1)\cos(\gamma)$. We therefore consider this interval
in the 
subsequent collisions and reduce our parameter space to the variables
$\gamma_n$ and $\theta_n$. Integrating $\rho_1(t_1,\gamma_1,\theta_1)$ over
$t_1$, we will get:
\begin{equation}
\rho_1(\gamma_1,\theta_1)=\frac{1}{\pi^2}\cos^2(\gamma_1).
\label{rho1gth}
\end{equation}
Now we may relate the variables at the first and second times the border is
reached:
\begin{subequations}
\label{transf2}
\begin{eqnarray}
\gamma_1 &=& \gamma_2,\\
\theta_1 &=& \theta_2-\pi+2\gamma_2.
\end{eqnarray}
\end{subequations}
The jacobian of this transformation is unitary, and thus the joint
distribution at the second time the border is reached will be
\begin{equation}
\rho_2(\gamma_2,\theta_2)=\frac{\cos^2(\gamma_2)}{\pi^2}
f(\theta_1,\delta),
\label{rho2}
\end{equation}
where the function $f(\theta_1,\delta)=1$ if $|\theta_1|>\delta$,
and vanishes otherwise, so that only particles which did not leave the vesicle
at the first time the border was reached do contribute. Actually $\theta_1$ in
the right hand side of equation \ref{rho2} should be written as a function of
$\theta_2$ and $\gamma_2$ using the equations \ref{transf2}. It is now easy to
generalize this result for the $n$'th time the particle reaches the border:
\begin{equation}
\rho_n(\gamma_n,\theta_n)=\frac{\cos^2(\gamma_n)}{\pi^2}\prod_{i=1}^{n-1}
f(\theta_i,\delta).
\label{jd}
\end{equation}
We remark that in the absence of the aperture, the joint distribution
will be the same at any collision, and therefore for the closed
circular billiard, the joint distribution \ref{rho1gth} and  the
marginal distributions \ref{g1} and \ref{th1} are the same for any
value of $n$. For $n>1$ and when the billiard is open, the joint
distribution \ref{jd} is not 
normalized, since particles leave the billiard through the aperture. 

Next, we may calculate the marginal distributions at the second time the
particle reaches the border. The results are
\begin{eqnarray}
\rho_2(\gamma_2) &=& \frac{\cos^2(\gamma_2)}{\pi^2} \int_{-\pi}^{\pi}
f(\theta_2-\pi+2\gamma_2,\delta) \; d\theta_2 \nonumber \\
&=& \frac{2}{\pi^2}(\pi-\delta)\cos^2(\gamma_2),
\end{eqnarray}
and
\begin{equation}
\rho_2(\theta_2)=\frac{1}{2\pi^2}[\pi-\delta+
\cos(\theta_2)\sin(\delta)].
\end{equation}
One notices that the distribution in $\theta_2$ is no longer uniform if
$\delta>0$, displaying a maximum at $\theta_2=0$. This may be explained
qualitatively by the effect that due to the aperture, it will be more probable
for a 
particle to have a positive than a negative horizontal component of the
velocity after the first collision, since all particles which left the vesicle
would have negative horizontal components if they had collided with the
wall. Also, comparing the marginal distributions $\rho_1(\gamma_1)$ and
$\rho_2(\gamma_2)$, we notice that they differ just by a factor
$1-\delta/\pi$, which accounts for the particles that left the vesicle
at the first time the border was reached. The distribution of $t^\prime_2$ may
be obtained from $\rho_2(\gamma_2)$: 
\begin{eqnarray}
\rho_2(t^\prime_2) &=& 2\rho_2(\gamma_2)\left|\frac{d \gamma_2}{d t^\prime_2}
\right| \nonumber \\
&=&\frac{2\tau_2^2(\pi-\delta)}{\pi^2\sqrt{1-\tau_2^2}},
\label{tp2}
\end{eqnarray}
where we defined $\tau_2=t^\prime_2/2$, which is restricted to the
interval $[0,2]$.
The probability that the particle
leaves the vesicle at the second time it reaches the border will be:
\begin{eqnarray}
p_2 &=& 2 \int_0^\delta \rho_2(\theta_2) d\theta_2 = \nonumber \\
&&\frac{1}{\pi^2}[\delta(\pi-\delta)+\sin^2(\delta)].
\label{p2}
\end{eqnarray}
As expected, this probability is larger than the value we would obtain for a
uniform distribution in $\theta_2$, which is $\delta(\pi-\delta)/\pi^2$.

In general, to calculate the marginal distributions at the $n$'th time the
particle reaches the border, we have to integrate the joint distribution
\ref{jd}. For the incidence angle we obtain:
\begin{eqnarray}
\rho_n(\gamma_n) &=& \frac{\cos^2(\gamma_n)}{\pi^2}\int_{-\pi}^\pi
\prod_{i=1}^{n-1} f(\theta_i,\delta) \; d\theta_n \nonumber \\
&=& \frac{\cos^2(\gamma_n)}{\pi^2} I_n(\gamma_n,\delta),
\label{gamman}
\end{eqnarray}
while the expression for the marginal distribution in $\theta_n$ is:
\begin{eqnarray}
\rho_n(\theta_n) &=& \frac{1}{\pi^2} \int_{-\pi/2}^{\pi/2} \cos^2(\gamma_n) 
\prod_{i=1}^{n-1} f(\theta_i,\delta) \; d\gamma_n \nonumber \\
&=& \frac{1}{\pi^2} J_n(\theta_n,\delta).
\label{thetan}
\end{eqnarray}
The distribution of the time $t^\prime_n$ may be found generalizing
equation \ref{tp2}. The result is:
\begin{equation}
\rho_n(t^{\prime}_n)=\frac{\tau_n^2\;I_n(\gamma_n,\delta)}
{(n-1)\pi^2\sqrt{1-\tau_n^2}},
\label{tpn}
\end{equation}
where $\tau_n=t^\prime_n/2(n-1)$ and $\cos(\gamma_n)=\tau_n$. It may
be useful to stress that all probability distributions are normalized
with respect to the initial condition, and thus the area under each
marginal distribution is an decreasing function of $n$, since particles are
leaving the vesicle at each time the border is reached. 

\begin{figure}
\begin{center}
\includegraphics[height=6.0cm]{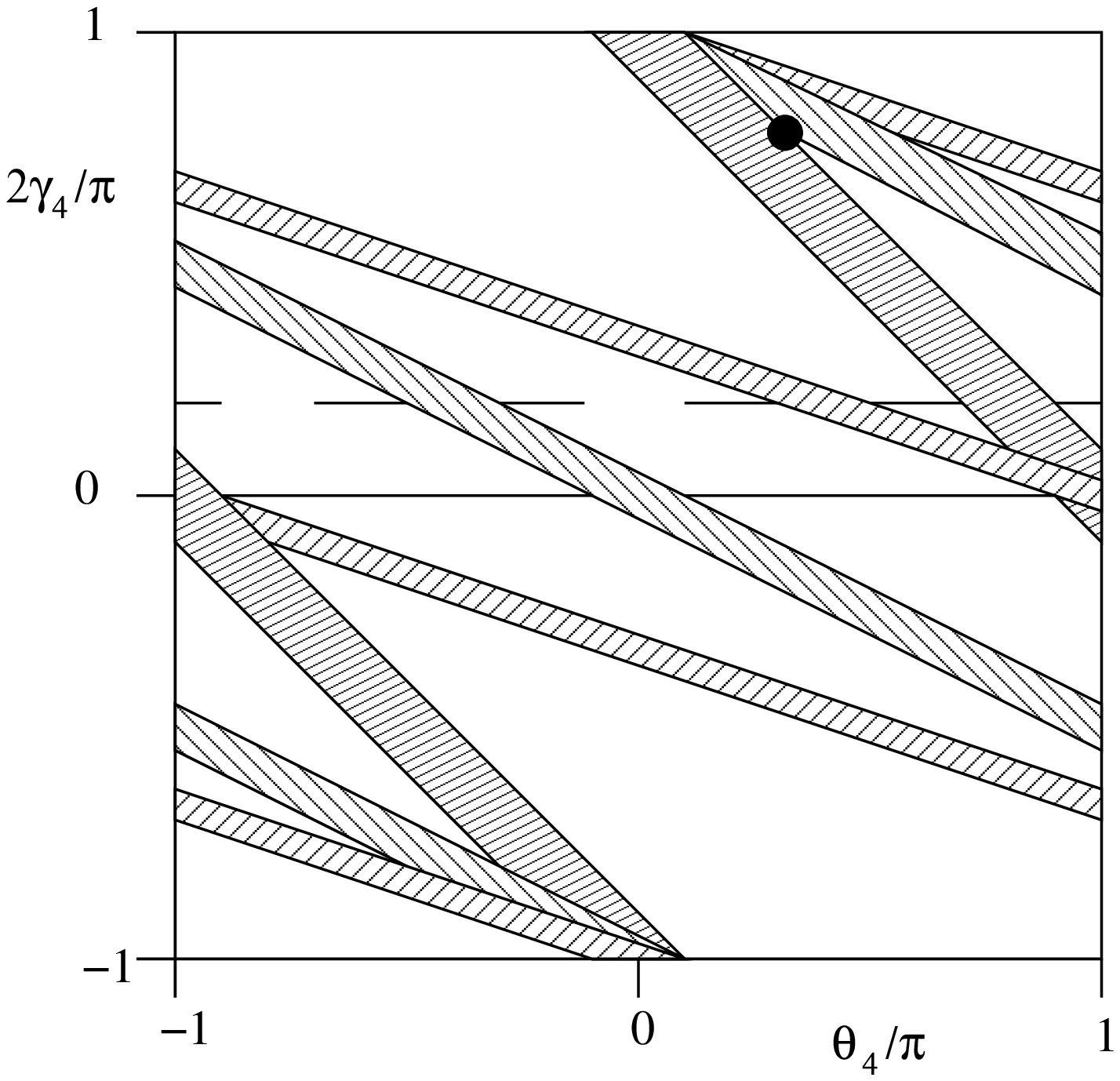}
\caption{Graphical representation of
 $\prod_{i=1}^{n-1} f(\theta_i,\delta)$, for $n=4$ and
 $\delta=\pi/10$ as a function of $\theta_4$ and $\gamma_4$. The product 
 vanishes in the shaded areas and is equal to one in the
 complement. The horizontal segments
 at $\gamma_1=\pi/10$ and $\gamma_1=0$ correspond to initial
 conditions of a five-point star and diagonal periodic orbits,
 respectively, as discussed in the text. Particles with these
 initial conditions never leave the vesicle. The circle indicates a
 particular crossing point of borders of stripes mentioned in the
 appendix.}
\label{f2}
\end{center}
\end{figure}

To calculate of the integrals $I_n$ and $J_n$ defined above, we have to
identify the region in the $(\theta_n,2\gamma_n)$ plane where at least one of
the $f$ functions in the product vanish. The region where a particular
function $f(\theta_{n-j},\delta)$ vanishes is a set of $j+1$ parallel
stripes, as may be seen in figure \ref{f2}. In this figure, which
corresponds to $n=4$, the two wider stripes correspond to
$(\theta_4$, $2\gamma_4)$ values such that $|\theta_3|<\delta$, the
three stripes of intermediate width correspond to $|\theta_2|<\delta$
and points inside the four narrowest stripes are such that
$|\theta_1|<\delta$. The expressions for the stripes are obtained by a
recursive application of equations \ref{transf2} which relate the
variables at successive times the particle reaches the border.

The integrals $I_n$ and $J_n$ are continuous functions of their
variables, but may display discontinuities in their derivatives. Their
expressions are different in distinct ranges of their
arguments. Although the calculation of these functions is rather
straightforward, the identification of the ranges of integration
becomes harder as the number of the collision $n$ grows, the number of
different expressions increases quite fast with $n$. In the general
case, we performed the calculations up to $n=4$. For $\delta \geq
\pi/2$, however, the calculation of the marginal distributions may be
accomplished for any value of $n$. Some details of these calculations and the
results are shown in the appendix.

Using the results for the marginal distribution $\rho_3$ we may obtain
the probabilities: 
\begin{eqnarray}
p_3 &=& \frac{1}{2\pi^2}[\delta(2\pi-3\delta)-1+2\cos(\delta)-
\cos(2\delta)],\nonumber \\
&&\mbox{for $0 \leq \delta \leq \frac{\pi}{2}$;}\label{p31}\\
p_3 &=& \frac{1}{2\pi^2}[\delta(\delta-2\pi)+\pi^2+1+\cos(2\delta)+
\nonumber \\
&&2\cos(\delta)],\mbox{ for $\frac{\pi}{2} \leq \delta \leq \pi$.}
\label{p32}
\end{eqnarray}
The probability that the particle leaves the vesicle at its
fourth arrival at the border is
\begin{eqnarray}
p_4 &=& \frac{1}{6\pi^2}\left[\delta(6\pi-13\delta)-12+
9\cos\left(\frac{2}{3}\delta\right)+6\cos(\delta)-\right.\nonumber \\
&&\left.3\cos(2\delta)\right], \mbox{for $0 \leq \delta \leq
    \frac{\pi}{3}$;}\label{p41}\\
p_4 &=& \frac{1}{6\pi^2}\left[\delta(5\delta-6\pi)+2\pi^2-3+
9\cos\left(\frac{2}{3}\delta\right)-\right.\nonumber \\
&&\left.6\cos(\delta)+3\cos(2\delta)
\right], \mbox{for $\frac{\pi}{3} \leq \delta \leq \frac{\pi}{2}$;} 
\label{p42}\\
p_4 &=& \frac{1}{12\pi^2}\left[2\delta(\delta-2\pi)+2\pi^2+6+
9\cos\left(\frac{2}{3}\delta\right)\right.-\nonumber \\
&&\left. 9\sqrt{3}\sin\left(\frac{2}{3}\delta\right)-12\cos(\delta)\right],
\mbox{for $\frac{\pi}{2} \leq \delta \leq \pi$.}\label{p43}
\end{eqnarray}
Finally, using the marginal density for general $n \geq 2$ obtained for
$\delta \geq \pi/2$ in the appendix, we obtain:
\begin{eqnarray}
p_{n+1} &=& \frac{1}{2\pi^2n(n-1)}\left[
  n(n-1)+2\delta^2+2\pi(\pi-2\delta)-\right.\nonumber \\
&& n^2(n-1)\cos\left(\frac{(n-3)\pi+2\delta}{n-1}\right)-\nonumber \\
&&n(n-1)\cos\left(\frac{2(\delta-\pi)}{n-1}\right)-\nonumber \\
&&\left. n^2(n-1)\cos\left(\frac{2(\delta-\pi)}{n}\right)\right].
\label{pn}
\end{eqnarray}

\section{Discussion of the results and comparison with simulations}
\label{simu}
\begin{figure}
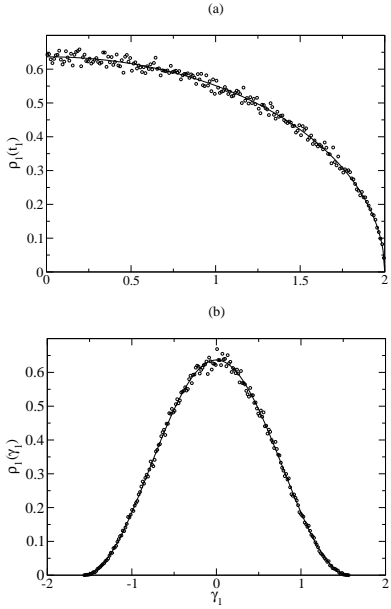

\begin{center}
\includegraphics[height=4.0cm]{rhot1.eps}
\includegraphics[height=4.0cm]{rhog1.eps}
\caption{Results from simulations for $\rho_1(t_1)$ (graph a) and
  $\rho_1(\gamma_1.)$ (graph b). The numerical results
  are indicated by circles and the full lines are the corresponding
  analytical curves (equations \ref{t1} and \ref{g1}, respectively).}
\label{f3}
\end{center}
\end{figure}

Initially, we show results for $\rho_1(t_1)$ obtained with a
simulation of $10^{11}$ particles starting at the initial condition,
that is, initial position inside the unitary circle with uniform
probability density and the direction of the initial velocity also
uniformly distributed. In figure \ref{f3} the simulational results for
the time of first arrival $t_1$ are shown and compared with the
analytical expression (\ref{t1}). In the same figure the numerical results
for $\rho_1(\gamma_1)$ are compared with expression (\ref{g1}). It is
apparent that the numerical results are compatible with the analytical
curves. 

Although we realized many simulations starting from the initial condition,
most of the numerical results shown here were obtained starting the
simulation at the first collision of the particle with the wall
($n=1$). With the same computational effort, these simulations led to
much better results than the ones starting from the initial
condition, since only two variables define this condition now, instead
of the three we have if the simulation starts at the initial
condition.The numerical procedure followed in these simulations for a 
particular value of $\delta$ was:
\begin{enumerate}
\item $N_\theta$ of values for $\theta_1$, uniformly spaced 
  in the interval $[-\pi,-\delta] \cup [\delta,\pi]$ are chosen. 
\item For each value of $\theta_1$, $N_\gamma$ values for the
  incidence angle $\gamma_1$ are generated randomly with the
  probability density \ref{g1}, and the subsequent loci of arrival of
  the particle are calculated for $n=2,3,\ldots$, up to a maximum
  number $n_{max}$. 
\item At the $n$'th arrival at the border, the values of $\theta_n$,
  $\gamma_n$, and $t^\prime_n$ are recorded, so that estimates for the
  marginal probability densities of these variables are obtained. If
  $-\delta \leq \theta_n \leq \delta$, the counter of the number of
  particles leaving the vesicle at the $n$'th arrival at the border is
  increased by one and the simulation of a new particle starts. If
  $n=n_{max}$, the simulation of a new particle is started, otherwise
  we proceed to step $n+1$. 
\end{enumerate}
The results shown below were obtained with the choices $N_{\theta}=10^5$,
$N_{\gamma}=10^5$, so that a total of $10^{10}$ particles are
considered in each simulation. We chose $n_{max}=30$.

\begin{figure}
\begin{center}
\includegraphics[height=6.0cm]{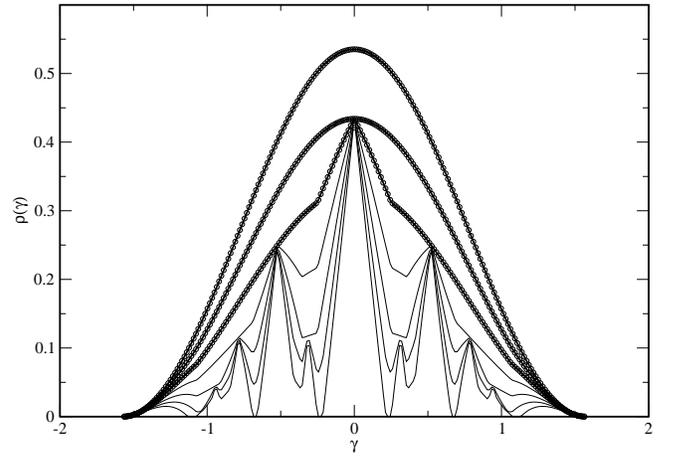}
\caption{Results for $\rho_n(\gamma_n)$ for $n=2,3,\ldots 8$. For the
  same value of $\gamma$, $\rho_n$ is a non-increasing function of
  $n$. For $n=2,3,4$ the numerical results 
  are indicated by circles and the full lines are the
  corresponding analytical curves (equation \ref{gamman}). For larger
  values of $n$, the lines correspond to results of
  simulations. Values of $n$ increase downwards.}
\label{f4}
\end{center}
\end{figure}

In figure \ref{f4} we show simulational and analytical results for
the marginal distribution $\rho(\gamma)$ when $\delta=0.5$.
One notices that the agreement between simulation and theory is very
good for the first four values of $n$, much better than the one
observed in figure \ref{f3}. This is due to the reduction of the
number of initial conditions mentioned above. Another feature which is
apparent in these data is that as $n$ grows, peaks appear in the
distribution at specific values of $\gamma$. The first peak to appear
is located at $\gamma=0$ and is already apparent in the analytical
expression for $\rho_4(\gamma_4)$. These peaks are associated to
periodic orbits in the vesicle, already discussed in
\cite{rr99}. Between two arrivals at the border,  
the angle $\theta$ changes as $\Delta \theta=\pi-2\gamma$. If the
orbit closes after $p$ collisions and the particle performs $q$ turns
in the vesicle, we have $p\Delta \theta=2\pi q$. Thus, the angle
$\gamma_{p,q}$ for this periodic orbit is:
\begin{equation}
\gamma_{p,q}=\frac{p-2q}{2p}\pi,
\end{equation}
where $p$ and $q$ have no common divisors, $p \ge 2q$ and we
considered only particles moving 
counterclockwise ($\Delta\theta>0$). The periodic orbits may therefore
be associated to the rational numbers $q/p$. For example, $p=2,q=1$,
corresponds to 
$\gamma=0$: the particle moves along a diameter. A polygonal orbit
with $p$ vertexes corresponds to $(p,1)$. A simple star with 5
vertices's is described by $(5,2)$. 

In an open vesicle, only those periodic orbits with $\pi/p>\delta$
will last, and the peak will develop when $n>p$. This may be simply
understood if we consider that a particle in a periodic orbit
characterized by an index $p$ will reach the border of the vesicle at
$p$ equally spaced points. If the angle between two fist neighbors is
larger that $2\delta$, for appropriate values of $\theta_1$ the
particle never reaches the aperture and therefore remains inside the
vesicle forever. Of course these orbits are a set of zero measure in
the initial conditions. The four peaks
visible in figure \ref{f4} at non-vanishing $\gamma$ may be identified
as $(5,2)$ ($\gamma=\pi/10$), $(3,1)$ ($\gamma=\pi/6$), $(4,1)$
($\gamma=\pi/4$), and $(5,1)$ ($\gamma=3\pi/10$). The hexagon $(6,1)$
still fulfils the condition above for the aperture $\delta=0.5$
($\pi/6>0.5$), but since it is quite close to the limit the
corresponding peak at $\gamma_{6,1}=\pi/3$ is very small and hard to
identify in the plots above. It is visible, however, in the results
for larger values of $n$. Once a peak appears, it will narrow with
increasing $n$. This may be explained noting that, of the particles
with the corresponding angle $\gamma_{p,q}$, some will leave the
vesicle, but only when $n \leq p$. Thus, the maximum value at the peak
is stationary for $n>p$. However, particles with a slightly different
incidence angle still leave the vesicle for $n>p$. As $n \to \infty$
the peak shrinks to just one point. The condition on the initial value
of $\theta$ which assures that a particle with $\gamma=\gamma_{p,q}$
will never leave the vesicle is $\ell |\theta_1|>\delta$, for
$\ell=1,2,\ldots p$ and reducing each angle to the interval
$[-\pi,\pi[$. In summary, the periodic orbits correspond to horizontal
segments in the $(\theta_1,\gamma_1)$ space of initial conditions. The
initial conditions for a periodic orbit labeled as $(q,p)$ in a
vesicle with aperture $\delta$ are $p$ segments at
$\gamma_1=\gamma_{p,q}$, with length $\Delta \Theta_1=2(\pi/p-\delta)$
and endpoints $|\theta|$ at
$(\delta+2\ell\pi/p,\delta+\ell\pi/p+\delta\theta)$, for
$\ell=0,1,2,\ldots,p/2$. In figure \ref{f2} these segments were drawn
for the star orbit $(5,2)$ and for the diagonal orbit $(2,1)$.

\begin{figure}
\begin{center}
\includegraphics[height=6.0cm]{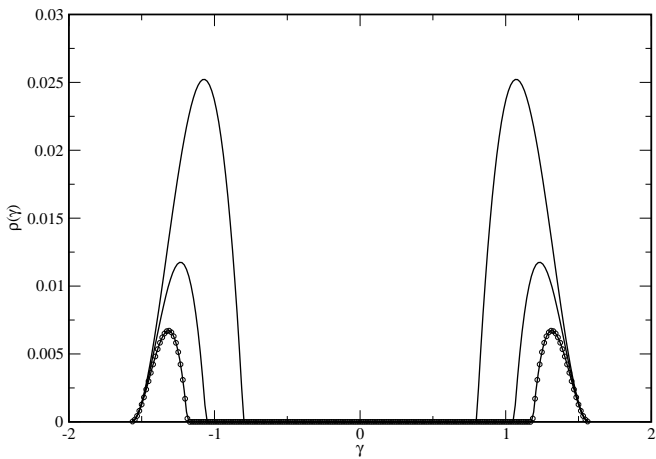}
\caption{Results for $\rho_n(\gamma_n)$ for $n=4,5,6$ and
  $\delta=1.6$. For $n=6$ the numerical results 
  are indicated by circles. Full lines are the
  analytical curves (equation \ref{gamman}).}
\label{f5}
\end{center}
\end{figure}

For $\delta>\pi/2$, all periodic orbits are suppressed, and we are
able to calculate the marginal distribution for any $n$. In figure
\ref{f5} results for $\delta=1.6$ are shown for $n=4,5,6$. The absence
of peaks related to any periodic orbit is apparent, and for $n=6$ the
agreement between the analytical result (line) and the simulations
(circles) may be appreciated. In this case, all particles with
$|\gamma|<((n+1)\delta-2\pi)/(n-1)$ leave the vesicle at the $n$'th
time the border is reached or before. 

\begin{figure}
\begin{center}
\includegraphics[height=6.0cm]{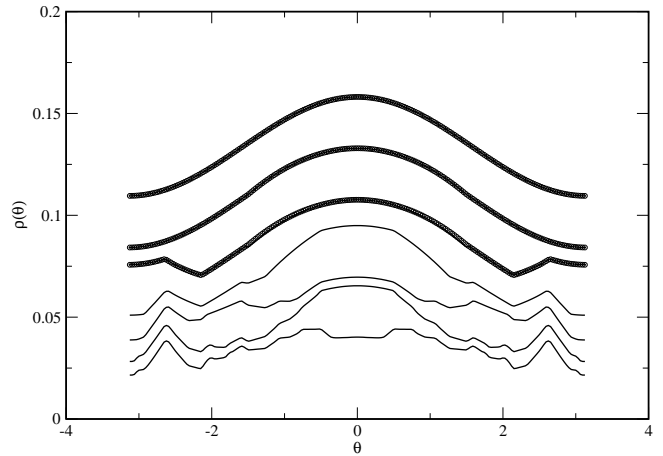}
\caption{Results for $\rho_n(\theta_n)$ for $n=2,3,4\ldots 8$ and
  $\delta=0.5$. For $n=2,3,4$ the numerical results 
  are indicated by circles and the full lines are the
  analytical curves (equation \ref{thetan}). For $n>4$, only
  simulational results are shown. Values of $n$ increase downwards.} 
\label{f6}
\end{center}
\end{figure}

Marginal distributions for $\theta$ are shown in figure \ref{f6} and
compared with the analytical results for $n \le 4$, also for the case
$\delta=0.5$. 
Again some peaks are observed, but they are much broader than the ones
in $\rho(\gamma)$. If we consider only a periodic orbit with $p$
vertexes, we would have $p$ equally spaced rectangular peaks, of width
$\Delta\theta_1$ defined above. Thus, as $p$ grows approaching the
limiting value $\pi/\delta$ the peaks become narrower. Therefore, for
$\delta=0.5$ the narrowest peaks belong to hexagons, and it is this
peak we may identify for largest values of $\theta$, located close to
$\theta=2.61$. For lower values of $\theta$ we have the superposition
of several peaks and identification is harder. However, for larger $n$
a step like structure develops at smaller $\theta$, as
expected. Although we do not show results here for $\delta>\pi/2$,
where we obtained $\rho_n(\theta_n)$ analytically for any $n$, as
expected smooth curves are found, since no periodic orbits survive
in this case. All particles with
\[
|\theta|<\frac{(n+1)\delta-2\pi}{n-1}
\]
leave the vesicle at the $n$'th arrival at the border or before.

\begin{figure}
\begin{center}
\includegraphics[height=6.0cm]{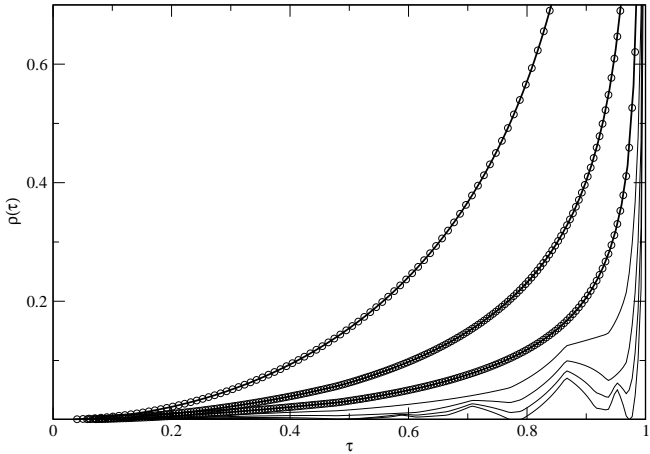}
\caption{Marginal density of probability of the time interval since
  the first arrival of the border, as a function of $\tau_n$, for
  $n=2,3,\ldots,8$ and $\delta=0.5$. Full lines for $n<5$ correspond
  to analytical results (equation \ref{tpn}), and the circles are
  results from the 
  simulations. For higher values of $n$ only numerical results are
  presented. Values of $n$ increase downwards.}
\label{f7}
\end{center}
\vspace{0.5cm}
\end{figure}

In figure \ref{f7} we show results for the marginal distributions in
the time $t^\prime_n$ between the $n$'th and first arrival of the
border. Again a good agreement between analytical end numerical
results is found for $n=2,3,4$. As expected, peaks develop at
values which correspond to surviving periodic orbits, as may already
be seen in the simulational results for larger values of $n$. The
largest peak located at $\tau_n=t^\prime_n/2(n-1)=1$ corresponds to
the diagonal periodic orbit $(2,1)$, the other periodic orbits
originate peaks at lower times. For $\delta>\pi/2$ again results for
general $n$ may be found and their agreement with simulational results
is very good. Since just polygonal paths $q=1$ survive after some 
collisions in this case, the maximum time a particle may spend inside
the vesicle is $2(\pi-\delta)$. This result may be obtained from the
general expression of $\rho_n(t^\prime_n)$ in the limit $n \to \infty$.  

\begin{figure}
\begin{center}
\includegraphics[height=6.0cm]{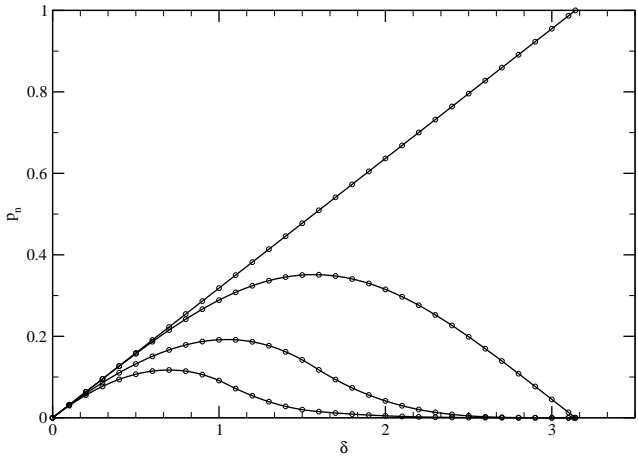}
\caption{Plots of $p_n$ as a function of $\delta$ for
  $n=1,2,3,4$. Results from simulations are represented by circles and
  the lines are analytic curves (equations \ref{p1}, \ref{p2},
  \ref{p31}-\ref{p32}, and \ref{p41}-\ref{p43}, respectively).}
\label{f8}
\end{center}
\end{figure}

\begin{figure}
\begin{center}
\includegraphics[height=5.8cm]{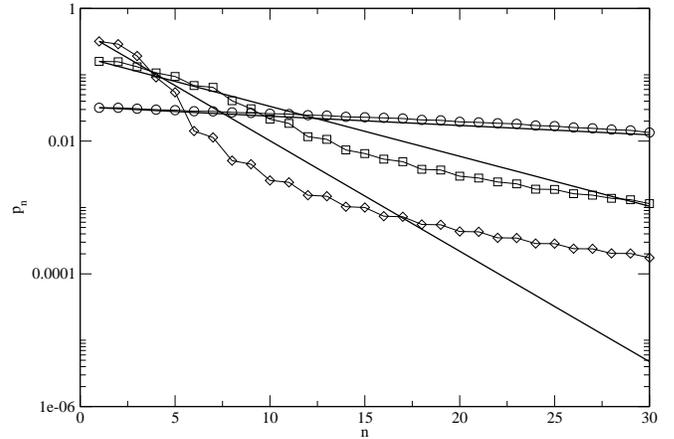}
\caption{Plots of $p_n$ as a function of $n$ for
  $\delta=0.1$ (circles), $\delta=0.5$ (squares), and $\delta=1.0$
  (diamonds). The thick lines are exponential functions discussed in
  the text.}
\label{f9}
\end{center}
\end{figure}

Figure \ref{f8} shows results for the probability $p_n$ that
a particle leaves the vesicle as the border is reached the $n$'th
time, as a function of the aperture $\delta$ for the cases where
analytic results are available. A good agreement is found between
numerical and analytical results.  
In figure \ref{f9} we present numerical results for $p_n$ as a
function of $n$ for some values of $\delta$. We also included
exponential functions which would be obtained for the escape
probability if the distribution $\rho(\theta)$ would be always
uniform, which are:
\begin{equation}
p_{n,u}=\left(1-\frac{\delta}{\pi}\right)^{n-1}\frac{\delta}{\pi}.
\end{equation}
First, we notice that the difference between the exponential decay and
the results is lower for small values of $\delta$. This may be
explained recalling that for the closed circular billiard $\delta=0$ the
distribution $\rho(\theta)$ is uniform, so that, at least for small
$n$, as $\delta$ increases the range of values of the probability
density becomes larger. As mentioned above, $p_2$ is always larger
than the value for an uniform distribution, but in general we observe
that for an intermediate range of $n$, the exponential values are
larger than the numerical results, but at sufficiently large $n$ they
again become smaller. For $\delta>\pi/2$, we may expand the expression
\ref{pn} in $1/n$ and find the asymptotic power law behavior:
\begin{equation}
p_{n+1}=\pi^2(1-\delta/\pi)^4\left[ \left(\frac{1}{n}\right)^4+2
\left(\frac{1}{n}\right)^5-\ldots\right],
\label{pna}
\end{equation} 
so that in this case we may assure that $p_n$ is larger than $p_{n,u}$
for sufficiently large values of $n$. It remains an open question if
this conclusion extends to $\delta<\pi/2$, but our numerical results
are consistent with this possibility. For $n=3,4$, $p_n$ is larger than
$p_{n,u}$ for $\delta$ in the range $]0,p_0[$ and smaller  in the range
$]p_0,\pi[$, where $p_0=\pi/2$ for $n=3$ and $p_0=0.919725\ldots$
for $n=4$.

\section{Conclusion}
\label{con}
In our study of the open circular billiard, it is natural to obtain
the distribution of the quantities of interest as a function of the
collision number $n$. In other approaches, as usual, the time appears
as the variable in terms of which the probability distributions are
calculated. Although we did discuss somewhat the distribution of times
at the first collisions, we did not enter in much detail in this question
here. This is the subject of ongoing work. In the simulations, we also
did not record results beyond 30 collisions with the wall, so that the
question of the asymptotic decay of $p_n$ for $\delta<\pi/2$ was not
discussed in detail. As stated above, we have indications of a power
law behavior, but the exponent may be different from the value 4 found
exactly for larger apertures. One may imagine that the existence of
orbits in which the particles are trapped forever inside the vesicle
for $\delta<\pi/2$ may lead to a slower decay, with exponents smaller
than 4 in this case. It is true, as may be seen above, that these
orbits are a set of zero measure in the initial conditions, but
particles in a finite region close to those orbits may take a long time
to exit. 

As was shown, the calculation of marginal distributions in the general
case becomes more difficult for increasing $n$ and finite
$\delta<\pi/2$. There is, however, a simplification when $\delta \ll
1$, that is, if we consider the first terms in the expansion of the
distributions in powers of $\delta$. This is the range where other
approaches, such as the one in \cite{vk01}, are effective. 

\begin{figure}
\begin{center}
\includegraphics[height=5.8cm]{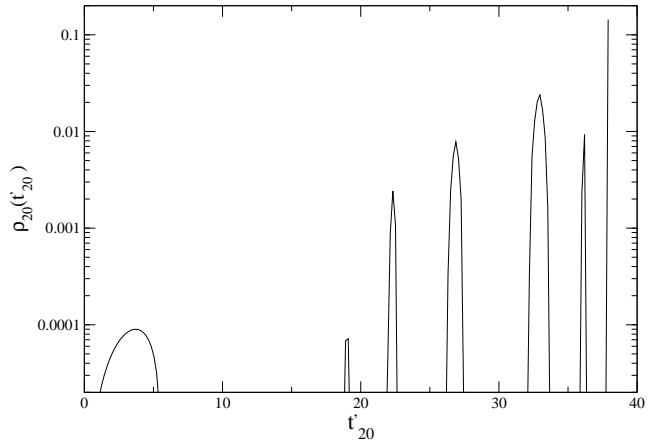}
\caption{Probability density of the time interval $t^\prime_{20}$
  obtained from simulations with $\delta=0.5$. The peaks which
  correspond to the neighborhood of surviving periodic orbits
  identified in the text are apparent. The broad peak at low times
  corresponds to whispering gallery modes still present in the
  vesicle.} 
\label{f10}
\end{center}
\end{figure}

There is a promising possibility to access the asymptotic
behavior at long times if we recall that, after a certain number of
collisions with the wall, the particles remaining in the vesicle will
have incidence angles close to the values defined for the surviving
periodical orbits. As another illustration of this, in figure
\ref{f10} numerical results for the distribution of the time interval
$t^\prime_{20}=t_{20}-t_1$ are depicted in a logarithmic scale for a
simulation with $\delta=0.5$. The most
important peak at $t^\prime_{20}=38$ corresponds to the vicinity of
the diagonal orbit $(2,1)$. The other peaks in order of decreasing
values of $t^\prime_{20}$ are the same mentioned in the discussion of
figure \ref{f4} above, but now the small peak corresponding to the
hexagon $(6,1)$ is visible centered at
$t^\prime_{20}=38\cos(\gamma_{6,1})=38\cos(\pi/3)=19$. The small
continuous distribution at low times corresponds to particles in high
$p$ orbits (whispering gallery modes), which still did not reach the
aperture after 19 
collisions. As discussed above, this distribution is located at times
smaller than $2(\pi-\delta)$ and does not contribute at larger
times. We are presently working on an approach where the neighborhood
of the periodic orbits is treated in a approximate way, which is
should be appropriate for the discussion of the long time limit.

\section*{Acknowledgements}

We thank Prof. Ronald Dickman for having brought reference \cite{vk01}
to our attention and Profs. Yan Levin, Marco Idiart, and Domingos
U. Marchetti for helpful 
discussions. This work was partially financed by project
Pronex-CNPq-FAPERJ/\-171.168-2003. 

\appendix

\section{Calculation of $I_n$ and $J_n$}

\begin{figure}
\begin{center}
\includegraphics[height=6.0cm]{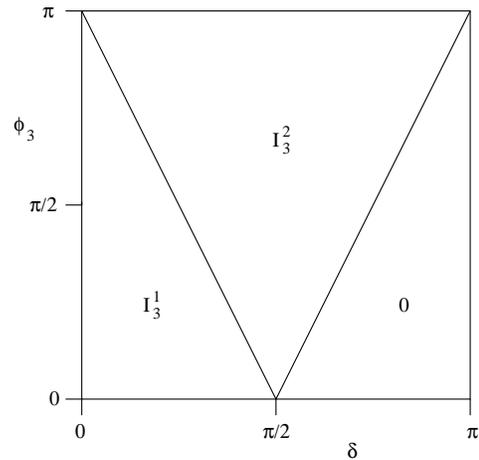}
\caption{Regions of the domain of the arguments where
  $I_3(\gamma_n,\delta)$ is given by different expressions. The
  function $I_3$ vanishes in the triangular region labeled by $0$.}
\label{fa1}
\end{center}
\end{figure}

Both $I_n(\gamma_n,\delta)$ and $J_n(\theta_n,\delta)$ are even
functions of their first argument, so we restrict the calculations
below to non-negative values of these variables. As may be seen in
figure \ref{f2}, the domains of integration are split into regions
where the product of $f$ functions is equal to unity. This regions
change as the variables of the marginal distributions are varied, so
that these distributions are given by different expression in
different areas of the domain of their arguments. It is convenient to
use the variable $\phi_n=2\gamma_n$ in the functions $I_n$. For $I_3$ these
areas are shown in figure \ref{fa1}.

For a particular value of the argument $\phi_3$, $I_3$ is equal to
to the sum of the segments in white regions of the corresponding
horizontal line in figure \ref{f2}, if we erase the narrowest stripes
which correspond to $n=4$. For $\delta \leq \pi/2$, this total length
is equal to $2\pi-4\delta$ if $\phi_3 \leq \pi-2\delta$ and to
$\pi+\phi_3-2\delta$ otherwise. In general, the expressions for $I_3$ in
the triangular regions depicted in figure \ref{f3} are
\begin{eqnarray}
I_3^1 &=& 2\pi-4\delta, \\
I_3^2 &=& \pi+\phi_3-2\delta.
\end{eqnarray}
In these calculations, the crossings of the borders of the stripes
determine the limits of the region in the arguments where each
expression is valid. These borders have the general form 
\begin{equation}
\theta_{n+1}+n\phi_{n+1}=(n-2\ell)\pi \pm \delta,
\end{equation}
where the plus sign is for the upper border and the minus sign for the
lower border of the stripe and $0 \leq \ell \leq n$ labels the stripes
in each order $n$. As an example of such a crossing of borders, we
have the point indicated by a circle  in figure \ref{f2}, which
corresponds to the crossing of the lines $(n=3,\ell=0,-)$ and
$(n=2,\ell=0,+)$ and is located at $\theta=3\delta$,
$\phi=\pi-2\delta$. 

\begin{figure}
\begin{center}
\includegraphics[height=6.0cm]{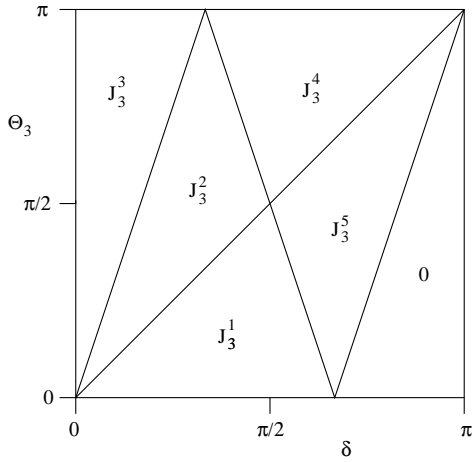}
\caption{Regions of the domain of the arguments where
  $J_3(\gamma_n,\delta)$ is given by different expressions. The
  equations of the limiting lines are $\theta_3=\delta$,
  $\theta_3=3\delta$, $\theta_3=2\pi-3\delta$, and
  $\theta_3=\delta-2\pi$. $J_3$ vanishes in the region labeled by 0.}
\label{fa2}
\end{center}
\end{figure}

For the calculation of $J_n$, each vertical segment in the
white region of the diagram \ref{f2},limited by $\phi_-$ and $\phi_+$
contributes with 
$g(\phi_+)-g(\phi_-)$ to the function $J_n$, where integrating equation
(\ref{thetan}) we get:
\begin{equation} 
g(\phi)=\frac{1}{4}[\phi+\sin(\phi)].
\end{equation}
We may then obtain the expressions for $J_3$. The diagram indicating
the region where each expression is valid is shown in the figure \ref{fa2}.
The expressions are:
\begin{eqnarray}
J_3^1 &=& \frac{1}{2}\left[ \pi-\frac{3}{2}\delta-
  \cos(\theta_3/2)\sin(\delta/2) +\right. \nonumber \\
  &&\left. \cos(\theta_3)\sin(\delta)\right],
\\
J_3^2 &=& \frac{1}{2}\left[ \pi-\frac{\theta_3}{4}-\frac{5}{4}\delta
  +\frac{1}{2}\sin(\theta_3+\delta)-\right.\nonumber \\
 &&\left. \frac{1}{2} \sin\left(
  \frac{\theta_3+\delta}{2} \right)\right],
\\
J_3^3 &=&
\frac{\pi}{2}-\delta+\frac{1}{2}\cos(\theta_3)\sin(\delta),
\\
J_3^4 &=& \frac{1}{2}\left[ \frac{\pi}{2}-\frac{\delta}{2}
  -\sin\left(\frac{\theta_3}{2}\right)
  \cos\left(\frac{\delta}{2}\right)\right],
\\
J_3^5 &=& \frac{1}{4}\left[ \pi+\frac{\theta_3}{2}-\frac{3}{2}\delta
  -\sin(\theta_3-\delta) \right. -\nonumber \\
  &&\left.\sin\left(\frac{\theta_3+\delta}{2}
  \right)\right].
\end{eqnarray}

\begin{figure}
\begin{center}
\includegraphics[height=6.0cm]{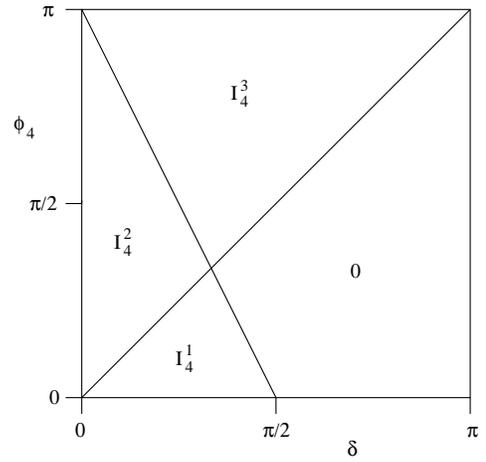}
\caption{Regions of the domain of the arguments where
  $I_4(\phi_4,\delta)$ is given by different expressions. The
  equations of the limiting lines are $\phi_4=\delta$ and
  $\phi_4=\pi-2\delta$. $I_4$ vanishes in the region labeled by 0.}
\label{fa3}
\end{center}
\end{figure}

For $n=4$, the function $I_4(\gamma_4,\delta)$ is given by three
different expressions in the triangular regions indicated in figure
\ref{fa3}, vanishing in the fourth region in the space of its
arguments. The expressions are:
\begin{eqnarray}
I_4^1 &=& 2(\pi-\phi_4-2\delta)\\
I_4^2 &=& 2(\pi-3\delta)\\
I_4^3 &=& 2(\phi_4-\delta).
\end{eqnarray}

\begin{figure}
\begin{center}
\includegraphics[height=6.0cm]{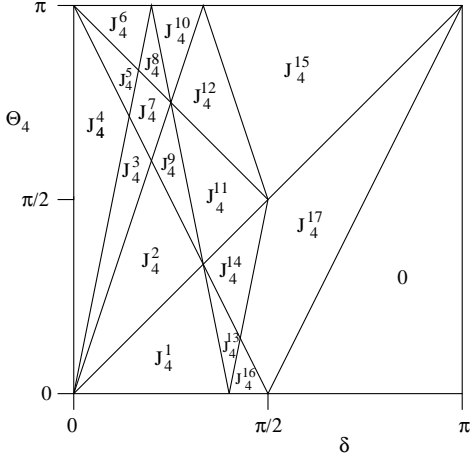}
\caption{Regions of the domain of the arguments where
  $J_4(\theta_4,\delta)$ is given by different expressions. The
  equations of the limiting lines are $\theta_4=\delta$,
  $\theta_4=3\delta$, $\theta_4=5\delta$, $\theta_4=2\pi-5\delta$,
  $\theta_4=\pi-2\delta$, $\theta_4=\pi-\delta$,
  $\theta_4=2\pi-3\delta$, $\theta_4=-\pi+2\delta$, and
  $\theta_4=5\delta-2\pi$. $J_4$ vanishes in the region labeled by 0.}
\label{fa4}
\end{center}
\end{figure}

Finally, in figure \ref{fa4} the domain of the arguments of the
function $J_4(\theta_4,\delta)$ is depicted, with a total of 18
regions where different expressions are valid.
The expressions are:
\begin{eqnarray}
J_4^1 &=& \frac{1}{2} \left[ \pi-\frac{13}{6}\delta+\cos(\theta_4)
  \sin(\delta)-\right. \nonumber \\
&&\left. \cos\left(\frac{\theta_4}{3}\right)
  \sin\left(\frac{\delta}{3}\right)-\right.\nonumber \\
&&\left. \cos\left(\frac{\theta_4}{2}\right)-
  \sin\left(\frac{\delta}{2}\right)\right],\\
J_4^2 &=& \frac{1}{2} \left[ \pi-\frac{\theta_4}{3}-\frac{11}{6}\delta
  +\frac{1}{2}\sin(\theta_4+\delta)-\right.\nonumber \\
&&\left. \frac{1}{2}\sin\left(\frac{\theta_4+\delta}{3}\right)-
  \cos\left(\frac{\theta_4}{2}\right) \sin
  \left(\frac{\delta}{2}\right)\right], \\
J_4^3 &=&  \frac{1}{2} \left[
  \pi-\frac{\theta_4}{12}-\frac{31}{12}\delta+\cos(\theta_4)\sin(\delta)
-\right.\nonumber\\
  &&\left. \frac{1}{2}\sin\left(\frac{\theta_4+\delta}{3}\right)+
  \frac{1}{2}\sin\left(\frac{\theta_4-\delta}{2}\right)\right],\\
J_4^4 &=&  \frac{1}{2} \left[ \pi-3\delta+\cos(\theta_4)\sin(\delta) 
  \right],\\
J_4^5 &=& \frac{\pi}{3}+\frac{\theta_4}{6}-\frac{7}{6}\delta-
  \frac{1}{4}\sin(\theta_4-\delta)+ \nonumber \\
&&\frac{\sqrt{3}}{8}\cos\left(\frac{\theta_4-\delta}{3}\right)
  -\frac{1}{8}\sin\left(\frac{\theta_4-\delta}{3}\right),\\
J_4^6 &=& \frac{\pi}{2}-\frac{4}{3}\delta+
  \frac{1}{2}\cos(\theta_4)\sin(\delta)+\nonumber \\
&&\frac{\sqrt{3}}{4}
  \sin\left(\frac{\theta_4}{3}\right)\sin\left(\frac{\delta}{3}
  \right)+\nonumber \\
&&\frac{1}{4}\cos\left(\frac{\theta_4}{3}\right)
  \sin\left(\frac{\delta}{3}\right),
\end{eqnarray}
\begin{eqnarray}
J_4^7 &=& \frac{\pi}{3}+\frac{\theta_4}{8}-\frac{23}{24}\delta+
  \frac{\sqrt{3}}{8}\cos\left(\frac{\theta_4-\delta}{3}\right)-\\
  \nonumber
&&\frac{1}{4}\sin\left(\frac{\theta_4+\delta}{3}\right)-
  \frac{1}{8}\sin\left(\frac{\theta_4-\delta}{3}\right)+
  \nonumber \\
&&\frac{1}{4}\sin\left(\frac{\theta_4-\delta}{2}\right)-
  \frac{1}{4}\sin(\theta_4-\delta),\\
J_4^8 &=& \frac{1}{2}\left[ \pi-\frac{\theta_4}{12}-\frac{9}{4}\delta+
  \cos(\theta_4)\sin(\delta)- \right. \nonumber \\
&&\frac{1}{2}\sin\left(\frac{\theta_4}{3}\right)
  \cos\left(\frac{\delta}{3}\right)+
  \frac{\sqrt{3}}{2}\sin\left(\frac{\theta_4}{3}\right) 
  \sin\left(\frac{\delta}{3}\right)+ \nonumber \\
&&\left. \frac{1}{2}\sin\left(\frac{\theta_4-\delta}{2}\right)\right],\\
J_4^9 &=& \frac{\pi}{3}+\frac{7}{12}\delta+
  \frac{\sqrt{3}}{8}\cos\left(\frac{\theta_4-\delta}{3}\right)-
  \nonumber \\
&&\frac{1}{4}\sin\left(\frac{\theta_4+\delta}{3}\right)-
  \frac{1}{8}\sin\left(\frac{\theta_4-\delta}{3}\right)-\nonumber \\
&&\frac{1}{2}\cos\left(\frac{\theta_4}{2}\right) 
  \sin\left(\frac{\delta}{2}\right),\\
J_4^{10} &=& \frac{1}{2}\left[ \frac{5}{6}\pi+\frac{11}{6}\delta-
  \frac{1}{2}\sin\left(\frac{\pi+\theta_4+\delta}{3}\right)+\right.
  \nonumber \\
&&\sin\left(\frac{\theta_4}{2}\right)\cos\left(\frac{\delta}{2}\right)+
  \cos(\theta_4)\sin(\delta)- \nonumber \\
&&\left.\frac{1}{2}\sin\left(\frac{\theta_4-\delta}{3}\right)\right],\\
J_4^{11} &=& \frac{1}{2}\left[\frac{\pi}{2}+\frac{\theta_4}{12}-
  \frac{3}{4}\delta-\sin\left(\frac{\theta_4}{3}\right)
  \cos\left(\frac{\delta}{3}\right)+ \right. \nonumber \\
&& \left. \frac{1}{2}\sin\left(\frac{\theta_4-\delta}{2}\right)
   \right], \\
J_4^{12} &=& \frac{1}{4}\left[\frac{5}{3}\pi+\frac{\theta_4}{2}-
  \frac{13}{6}\delta-
  \sin\left(\frac{\pi+\theta_4+\delta}{3}\right)+ \right. \nonumber \\
&&\sin\left(\frac{\theta_4-\delta}{2}\right)-
  \sin\left(\frac{\theta_4-\delta}{3}\right)+ \nonumber \\
&&\left. \sin(\theta_4+\delta) \right],\\
J_4^{13} &=& \frac{1}{2}\left[\frac{5}{6}\pi+\frac{1}{12}\theta_4-
  \frac{7}{4}\delta+\cos(\theta_4)\sin(\delta)- \right. \nonumber \\
&&\frac{1}{2}\sin\left(\frac{\pi+\theta_4+\delta}{3}\right)+
  \frac{1}{2}\sin\left(\frac{\theta_4-\delta}{2}\right)-
  \nonumber \\
&&\left. \sin\left(\frac{\theta_4+2\delta}{6}\right)
  \sin\left(\frac{\delta}{3}\right)\right],\\
J_4^{14} &=& \frac{1}{4}\left[\pi+\frac{5}{6}\theta_4-
  \frac{16}{6}\delta-\sin(\theta_4-\delta)- \right. \nonumber \\
&&\sin\left(\frac{\pi+\theta_4+\delta}{3}\right)+
  \sin\left(\frac{\theta_4-\delta}{2}\right)+\nonumber \\
&&\left.\cos\left(\frac{\pi+2\theta_4+2\delta}{6}\right)\right],\\
J_4^{15} &=& \frac{1}{2}\left[\frac{\pi}{3}-\frac{\delta}{3}-
  \frac{1}{2}\sin\left(\frac{\pi+\theta_4+\delta}{3}\right)- \right.
  \nonumber \\
&&\left. \frac{1}{2}\sin\left(\frac{\theta_4-\delta}{3}\right)\right],
\end{eqnarray}
\begin{eqnarray}
J_4^{16} &=& \frac{\pi}{3}-\frac{2}{3}\delta+
  \frac{1}{2}\cos(\theta_4)\sin(\delta)- \nonumber \\
&&\frac{\sqrt{3}}{4}\cos\left(\frac{\theta_4}{3}\right) 
  \cos\left(\frac{\delta}{3}\right)- \nonumber \\
&&\frac{1}{4}
  \cos\left(\frac{\theta_4}{3}\right)\sin\left(\frac{\delta}{3}\right),\\
J_4^{17} &=& \frac{\pi}{6}+\frac{\theta_4}{6}-\frac{\delta}{3}-
  \frac{1}{4}\sin(\theta_4-\delta)- \nonumber \\
&&\frac{1}{4}\sin\left(\frac{\pi+\theta_4+\delta}{3}\right).
\end{eqnarray}

The extension of these calculations to larger values of $n$ becomes
increasingly difficult and will not be presented here. However, for $n
\geq 2$ and $\delta \geq \pi/2$ a major simplification occurs and it
is easy to perform the calculations of $I_{n+1}$ and $J_{n+1}$,
since in the diagram corresponding to figure \ref{f2} in this case the
white area is reduced to only two triangles and two quadrangles. The
vertexes of the first triangle are located at $\theta=\phi=-\pi$,
$\theta=-\pi,\phi=-\pi(1-1/n)$, and $\theta=-\delta,\phi=-\pi$. The
second triangle may be obtained by transforming $\theta \to -\theta$
and $\phi \to -\phi$. The vertexes of the quadrangle are located at
$\theta=\pi,\phi=-\pi$, $\theta=\pi,\phi=-\pi(1-1/n)$,
$\theta=\delta,\phi=-\pi$, and
$\theta=(\delta(n+1)-2\pi)/(n-1),\phi=-((n-3)\pi+2\delta)/(n-1)$. The
second quadrangle may be obtained by the same projection operation
described above. We then obtain
\begin{eqnarray}
I_{n+1}&=&0,\mbox{for $0 \leq \phi_{n+1} \leq
  \frac{(n-3)\pi+2\delta}{n-1}$;}\\
I_{n+1}&=&(3-n)\pi-2\delta+(n-1)\phi_{n+1}, \nonumber \\
&&\mbox{ otherwise.}
\end{eqnarray}

For the function related to the marginal distribution in
$\theta_{n+1}$ we get the result $J_{n+1}=0$ for $0 \leq \theta_{n+1}
\leq ((n+1)\delta-2\pi)/(n-1)$ and:
\begin{eqnarray}
J_{n+1} &=& \frac{1}{4}\left[ \frac{2\pi-\theta_{n+1}-\delta}{n}+
\theta_{n+1}-\delta- \right.\nonumber \\
&&\sin\left(\frac{(n-2)\pi+\theta_{n+1}=\delta}{n}\right)-\nonumber\\
&&\left.\sin(\theta_{n+1}-\delta)\right], \nonumber \\
&&\mbox{for $\frac{(n+1)\delta-2\pi}{n-1} \leq \theta_{n+1} \leq \delta$;}\\
J_{n+1} &=& \frac{1}{4}\left[\frac{2}{n}(\pi-\theta_{n+1})\right.-
\nonumber \\
&&\sin\left(\frac{(n-2)\pi+\theta_{n+1}+\delta}{n}\right)- \nonumber \\
&&\left.\sin\left(\frac{\theta_{n+1}}{n}\right)\right],
\mbox{for $\delta \leq \theta_{n+1} \leq \pi$.}
\end{eqnarray}

\end{document}